\def\mtx{$M$-$T_{\rm X}$}

\def\tx{T_{\rm X}}
\def\tew{T_{\rm EW}}
\def\tsl{T_{\rm SL}}

\def\Rap{R_{\rm ap}}

\documentclass{emulateapj}
\begin{document}
\title{Substructure and Scatter in the Mass--Temperature Relations of
  Simulated Clusters}

\author{
  David A. Ventimiglia\altaffilmark{1},
  G. Mark Voit\altaffilmark{1},
  Megan Donahue\altaffilmark{1},
  S. Ameglio\altaffilmark{2}
}


\altaffiltext{1}{Michigan State University, Physics \& Astronomy
  Dept., East Lansing, MI 48824-2320; ventimig@msu.edu,
  voit@pa.msu.edu, donahue@pa.msu.edu}
\altaffiltext{2}{Dipartimento di Astronomia dell'Universit\`a di
  Trieste, via Tiepolo 11, I-34131 Trieste, Italy; ameglio@ts.astro.it}

\begin{abstract}
  Galaxy clusters exhibit regular scaling relations among their bulk
  properties.  These relations establish vital links between halo mass
  and cluster observables.  Precision cosmology studies that depend on
  these links benefit from a better understanding of scatter in the
  mass-observable scaling relations.  Here we study the role of merger
  processes in introducing scatter into the \mtx\ relation, using a
  sample of 121 galaxy clusters simulated with radiative cooling and
  supernova feedback, along with three statistics previously proposed
  to measure X-ray surface brightness substructure.  These are the
  centroid variation ($w$), the axial ratio ($\eta$), and the power
  ratios ($P_{20}$ and $P_{30}$).  We find that in this set of simulated clusters,
  each substructure measure is correlated with a cluster's departures
  $\delta \ln \tx$ and $\delta \ln M$ from the mean \mtx\ relation,
  both for emission-weighted temperatures $\tew$ and for
  spectroscopic-like temperatures $\tsl$, in the sense that clusters
  with more substructure tend to be cooler at a given halo mass.  In
  all cases, a three-parameter fit to the \mtx\ relation that includes
  substructure information has less scatter than a two-parameter fit
  to the basic \mtx\ relation.
\end{abstract}

\keywords{galaxies: clusters: general, X-rays: galaxies: clusters}

\section{Introduction}

Clusters of galaxies play a critical role in our understanding of the
Universe and its history and are potentially powerful tools for
conducting precision cosmology.  For example, large cluster surveys
can discriminate between cosmological models with different
dark-energy equations of state by providing complementary observations
of the shape of the cluster mass function, evolution in the number
density of clusters with redshift, and bias in the spatial
distribution of clusters
\citep{1998ApJ...508..483W, 2002ApJ...577..569L,
2003PhRvD..67h1304H,2004ApJ...613...41M,Voit:2005PhysRevD}.
However, this potential to put tight constraints on the properties of
dark energy will be realized only if we can accurately measure the
masses of clusters and can precisely characterize the scatter in our
mass measurements.

Scatter in X-ray cluster properties is directly related to
substructure in the intracluster medium.  If clusters were all
structurally similar, then there would be a one-to-one relationship
between halo mass and any given observable property.  Generally
speaking, deviations from a mean mass-observable relationship are
attributed to structural differences among clusters.  One kind of
structural difference is the presence or absence of a cool core, in
which the central cooling time is less than the Hubble time at the
cluster's redshift, and the prominence of a cool core is observed to
be a source of scatter in scaling relationships
\citep{1994MNRAS.267..779F,1998ApJ...504...27M,2002ApJ...576..601V,2004ApJ...613..811M}.
We also expect structural differences to arise from substructure in
the dark matter, galaxy, and gas distributions.  For instance, there
may be a spread in halo concentration at a given mass, variations in
the incidence of gas clumps, differences in the level of AGN feedback,
or various effects due to mergers.  All of these deviations can be
considered forms of substructure that produce scatter in the
mass-observable relations one would like to use for cosmological
purposes.  While it may ultimately be possible to constrain the amount
of scatter and its evolution with redshift using self-calibration
techniques \citep{2005PhRvD..72d3006L}, such constraints would be
improved by prior knowledge about the relationship between scatter and
substructure.

Traditionally, the most worrisome form of substructure has been that
due to the effects of merger events.  Clusters are often identified as
``relaxed'' or ``unrelaxed'', with the former assumed to be nearly in
hydrostatic equilibrium and the latter suspected of being far from
equilibrium.  Cosmological simulations of clusters indicate that the
truth is somewhere in between.  The cluster population as a whole
appears to follow well-defined virial relations with log-normal
scatter around the mean, showing that clusters do not cleanly separate
into relaxed and unrelaxed systems (e.g.,\cite{2007astro.ph..2241E}).
Even the most relaxed-looking clusters are not quite in hydrostatic
equilibrium (e.g.,\cite{2006ApJ...650..128K}).  Instead of simply
being ``relaxed'' or ``unrelaxed,'' clusters occupy a continuum of
relaxation levels determined by their recent mass-accretion history.

Quantifying this continuum of relaxation offers opportunities for
reducing scatter in the mass-observable relations.  If mergers are
indeed responsible for much of the observed scatter around a given
scaling relation, then there may be correlations between a cluster's
morphology and its degree of deviation from the mean relation.  Once
one identifies a morphological parameter that correlates with the
degree of deviation, one can construct a new mass-observable relation
with less scatter by including the morphological parameter in the
relation.  Such an approach would be analogous to the improvement of
Type Ia supernovae as distance indicators by using light-curve stretch
as a second parameter to indicate the supernova's luminosity
\citep{1993ApJ...413L.105P,1996ApJ...473...88R}.

Here we investigate how merger-related substructure in simulated
clusters affects the relationship between a simulated cluster's mass
and the temperature of its intracluster medium, building upon
\cite{1995ApJ...452..522B} and \cite{2006ApJ...639...64O}.
\cite{1995ApJ...452..522B} quantified the morphologies and dynamical
states of observed clusters and found structure measures to be an
indicator of the dynamical state of a cluster.
\cite{2006ApJ...639...64O} also examined morphological measurements,
for both observed and simulated clusters, and found that simulations
without cooling showed no correlation between substructure and scaling
relation scatter.  In this work we examine substructure for simulated
clusters with radiative cooling and focus on the idea that merger
processes introduce intrinsic scatter into the \mtx\ relationship by
displacing clusters in the \mtx\ plane away from the mean X-ray
temperature $\langle{\tx}\rangle|_{_M}$ at a given mass $M$, either to
higher or lower average ICM temperature.  We then adopt a set of
statistics \citep{1995ApJ...452..522B,2006ApJ...639...64O} for
quantifying galaxy cluster substructure and merger activity in order
to investigate this hypothesis.  Section 2 discusses the \mtx\ scaling
relationship in our sample of simulated clusters and shows that
disrupted-looking clusters in this sample tend to be cooler at a given
cluster mass.  In Section 3 we attempt to quantify the relationship
between morphology and temperature using four different substructure
statistics and compare it to similar studies.  We then show that
substructure in these simulated clusters indeed correlates with
scatter in the \mtx\ relationship and assess the prospects for using
that correlation to reduce scatter in the \mtx\ plane.  Section 4
summarizes our results.

\section{Mass-Temperature Relation in Simulated Clusters}

This study is based on an analysis of 121 clusters simulated using the
cosmological hydrodynamics TREE+SPH code GADGET-2
\citep{2005MNRAS.364.1105S}, which were simulated in a standard
$\Lambda$ cold dark matter ($\Lambda$CDM) universe with matter density
$\Omega_M$ = 0.3, $h$ = 0.7, $\Omega_b$ = 0.04, and $\sigma_8$ = 0.8.
The simulation treats radiative cooling with an optically-thin gas of
primordial composition, includes a time-dependent UV background from a
population of quasars, and handles star formation and supernova
feedback using a two-phase fluid model with cold star-forming clouds
embedded in a hot medium.  All but four of the clusters are from the
simulation described in \cite{2004MNRAS.348.1078B}, who simulated a
box $192 \, h^{-1} \, {\rm Mpc}$ on a side, with $480^3$ dark matter
particles and an equal number of gas particles.  The present analysis
considers the 117 most massive clusters within this box at $z = 0$,
which all have $M_{200}$ greater than $5 \times 10^{13} \, h^{-1}
M_\odot$.  By convention, $M_{\Delta}$ refers to the mass contained in
a sphere which has a mean density of $\Delta$ times the critical
density $\rho_c$, and whose radius is denoted by $R_{\Delta}$

That cluster set covers the $\sim$1.5-5 keV temperature range, but the
$192 \, h^{-1} \, {\rm Mpc}$ box is too small to contain significantly
hotter clusters.  We therefore supplemented it with four clusters with
masses $> 10^{15} \, h^{-1} \, M_\odot$ and temperatures $> 5$ keV
drawn from a dark-matter-only simulation in a larger $479 \, h^{-1} \,
{\rm Mpc}$ box \citep{2006MNRAS.373..397S}.  The cosmology for this
simulation also was $\Lambda$CDM, but with $\sigma_8$ = 0.9.  These
were then re-simulated including hydrodynamics, radiative cooling, and
star formation, again with GADGET-2 and using the
zoomed-initial-conditions technique of \cite{1997MNRAS.290..411T},
with a fourfold increase in resolution.  This is comparable to the
resolution of the clusters in the smaller box.  Adding these four
massive clusters to our sample gives a total of 121 clusters with
$M_{200}$ in the interval $5 \times 10^{13} \, h^{-1} M_\odot$ to $2
\times 10^{15} \, h^{-1} M_\odot$.

We first need to specify our definitions for mass and temperature.  In
this paper, cluster mass refers to $M_{200}$.  For temperature, we use
two definitions.  The first is the emission-weighted temperature
\begin{equation}
  \tew = \frac { \int T [n^2 \Lambda(T) ] \, d^3 x } { \int n^2 \Lambda(T) \, d^3 x }  \; \; ,
\end{equation}
where $n$ is the electron number density and $\Lambda(T)$ is the usual
cooling function for intracluster plasma.  The second is the
spectroscopic-like temperature of \cite{2004astro.ph..4425M},
\begin{equation}
  \tsl = \frac { \int T [n^2 T^{-3/4}]  \, d^3 x } { \int n^2 T^{-3/4} \,  d^3 x }  \; \; ,
\end{equation}
where the power-law weighting function replacing $\Lambda(T)$ is
chosen so that $\tsl$ approximates as closely as possible the
temperature that would be determined from fitting a single-temperature
plasma emission model to the integrated spectrum of the intracluster
medium.  The presence of metals in the ICM of real clusters introduces
line emission that complicates the computation of $\tsl$ for clusters
$<$3 keV \citep{2006ApJ...640..710V}.  However, the simulated spectra
for the clusters in our sample are modelled with zero metallicity,
which eases this restriction in our analysis.

\begin{figure}[t]
  \centering
  \includegraphics[width=0.5\textwidth]{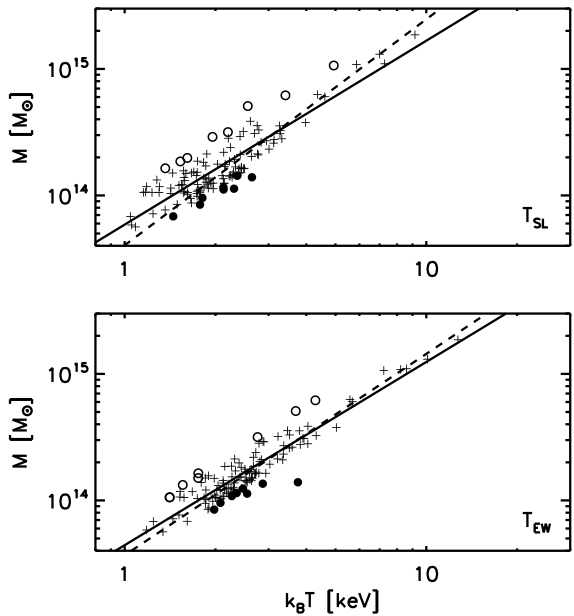}
  \caption{Mass-temperature (\mtx) relationships for the 121 clusters
    in our sample.  Spectroscopic-like temperature $\tsl$ is plotted
    in the top panel, while emission-weighted temperature $\tew$ is
    plotted in the bottom panel. The solid lines show the best-fit
    power-law relation $M = M_0 (\tx/3\ {\rm keV})^{\alpha}$ over the
    whole sample, while the dashed lines show the best-fit relation
    for systems with $\tx>$ 2 keV.  Open circles represent the
    clusters that have the greatest positive temperature offset from
    the mean relationship, and filled circles represent the clusters
    with the greatest negative temperature offset.}
  \label{fig:mass-temp}
\end{figure}

Figure \ref{fig:mass-temp} shows the mass-temperature relations based
on these definitions for our sample of simulated clusters.  The best
fits to the power-law form
\begin{equation}
  M = M_0 \left( \frac {\tx} {3 \, {\rm keV}} \right)^\alpha
  \label{eq:mass-temp}
\end{equation}
have the coefficients $M_0 \simeq 5.9 \times 10^{13}\ h^{-1}
M_{\odot}, \alpha \simeq 1.5$ for $\tx$ corresponding to $\tsl$ and
$M_0 \simeq 4.4 \times 10^{13}\ h^{-1} M_{\odot}, \alpha \simeq 1.4$
for $\tx$ corresponding to $\tew$.  As is generally the case for
simulated clusters, the power-law indices of the mass-temperature
relations found here are consistent with cluster self-similarity and
the virial theorem \citep{1986MNRAS.222..323K,1995MNRAS.275..720N}.
These relationships have scatter, which we characterize by the
standard deviation in log space $\sigma_{\ln M}$ about the best-fit
mass at fixed temperature $\tx$.  When relating $M$ to the
emission-weighted temperature $\tew$, we find $\sigma_{\ln M} =
0.102$.  When relating cluster mass $M$ to the spectroscopic-like
temperature $\tsl$, the scatter is $\sigma_{\ln M} = 0.127$.  That the
scatter is larger for the spectroscopic-like temperature is not
surprising, given the sensitivity of $\tsl$ to the thermal complexity
of the ICM.

Figure \ref{fig:mass-temp} also highlights two subsamples for each
definition of temperature, selected based on the clusters' deviations
in $\ln \tx$ space from the mean mass-temperature relation.  In each
panel, open circles represent the eight clusters that have the largest
positive deviations and are therefore ``hotter'' than other clusters
of the same mass, while filled circles represent the eight with the
largest negative deviation and are ``cooler'' than other clusters of
the same mass.  In general, these temperature estimates are well
correlated, so that hotter outliers in $\tew$ are also hotter outliers
in $\tsl$ and cooler outliers in $\tew$ are also cooler outliers in
$\tsl$.  Since Figure \ref{fig:mass-temp} distinguishes the most
extreme outliers for the two temperature estimates, this distinction
may define slightly different sets, though they still overlap.

\begin{figure}[t]
  \centering
  \includegraphics[width=0.5\textwidth,trim=0mm 40mm 0mm
  40mm,clip]{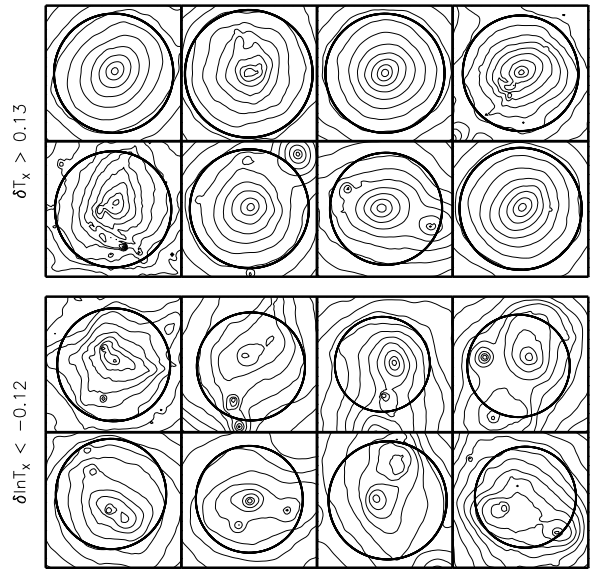}
  \caption{Surface-brightness contour maps for sixteen of the clusters
    in our sample, overlaid with a circular aperture of radius
    $R_{500}$.  The top panel has eight maps representing the clusters
    that have the largest positive deviation $\delta \ln \tx$ from the
    mean \mtx\ relation, calculated using the spectroscopic-like
    temperature $\tsl$.  These are the objects represented by open
    circles in the top panel of Figure \ref{fig:mass-temp}.  The
    bottom panel has eight maps representing the clusters that have
    the largest negative deviation from the same \mtx\ relationship,
    represented by the filled circles in Figure \ref{fig:mass-temp}.
    Note that the ``hotter'' clusters in the top panel appear more
    symmetric, while the ``cooler'' clusters in the bottom panel
    appear more irregular.}
  \label{fig:cluster_gallery_extremes}
\end{figure}

Figure \ref{fig:cluster_gallery_extremes} presents a gallery of
surface brightness maps for two sets of eight clusters with the most
extreme offsets from the mean $M$-$\tsl$ relation.  The eight
unusually hot clusters are in the top panel, and the eight cooler
clusters are in the bottom panel.  In these plots the hotter clusters
appear more symmetric, and are seemingly ``more relaxed,'' and the
cooler clusters appear less symmetric and seemingly ``less relaxed.''
The gallery as a whole therefore suggests that relaxed clusters tend
to be hot for their mass and unrelaxed clusters tend to be cool for
their mass.

At first glance, the result that disrupted-looking clusters in
cosmological simulations tend to be cooler than other clusters of the
same mass may seem counterintuitive, since one might expect that
mergers ought to produce shocks that raise the mean temperature of the
intracluster medium.  This finding has also been noted by
\cite{2001ApJ...546..100M} and \cite{2006ApJ...650..128K}.  Cluster
systems in the process of merging tend to be cool for their total halo
mass because much of the kinetic energy of the merger has not yet been
thermalized.

The idealized simulations of \cite{2007MNRAS.380..437P} illustrate
what may happen to the ICM temperature during a single merger.  Before
the cores of the two merging systems collide, the mean temperature is
cool for the overall halo mass because it is still approximately equal
to the pre-merger temperature of the two individual merging halos.
There is a brief upward spike in temperature when the cores of the
merging halos collide, after which the system is again cool for its
mass.  Then, as the remaining kinetic energy of the merger thermalizes
over a period of a few billion years, the temperature gradually rises
to its equilibrium value.  The merging system therefore spends a
considerably longer time at relatively cool values of mean temperature
for its halo mass than at relatively hot values.  Hence, such
simulations suggest a possible explanation for why more relaxed
systems would tend to lie on the hot side of the \mtx\ relation, while
disrupted systems would tend to lie on the cool side.  A caveat,
however, is that the current generation of hydrodynamic cluster
simulations tend to produce relaxed clusters whose temperature
profiles continue to rise to smaller radii than is observed in real
clusters \citep{2003MNRAS.342.1025T, 2007ApJ...668....1N}, potentially
enhancing average temperatures for such systems.  As a separate test
of this effect, we excise the core regions from our sample clusters,
calculate new substructure measures and new emission-weighted
temperatures for the core-excised clusters, and repeat our analysis.

\section{Quantifying Substructure}

The question we would like to address in this study is whether the
surface-brightness substructure evident in Figure
\ref{fig:cluster_gallery_extremes} is well enough correlated with
deviations from the mean mass-temperature relation to yield useful
corrections to that relation.  In order to answer that question, we
need to quantify the surface-brightness substructure in each cluster
image, so that we can determine the degree of correlation across the
entire sample.  \cite{2006ApJ...639...64O} explored the relationship
between cluster structure and X-ray scaling relations in both observed
and simulated clusters, and we adopt their suite of substructure
measures in this study.  These include centroid variation, axial
ratio, and the power ratios of \cite{1995ApJ...452..522B}.  In this
section we define and discuss those statistics and apply them to
surface-brightness maps made from three orthogonal projections of each
cluster.  Then we assess how well these statistics correlate with
offsets from the mean mass-temperature relation.

\subsection{Axial Ratio}
\begin{figure}[t]
  \centering
  \includegraphics[width=0.4\textwidth,trim=20mm 10mm 5mm 10mm,
  clip]{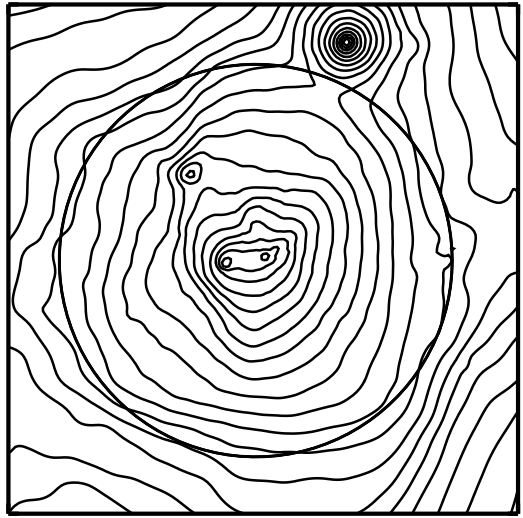}
  \caption{Surface-brightness contour plot of an asymmetric cluster
    which, for certain choices of aperture placement, yields an axial
    ratio close to 1.  The circle represents an aperture of
    $R_{500}$}.
  \label{fig:badaxialratio}
\end{figure}

The axial ratio $\eta$ for a cluster surface-brightness map is a
measure of its elongation, which is of interest because it has been
found from simulations that the ICM is often highly elongated during
merger events \citep{1993ApJ...419L...9E,1994MNRAS.268..953P}.  It is
computed from the second moments of the surface brightness,
\begin{equation}
  \label{eq:aratio_moments}
  M_{ij}=\sum I_{\rm X} x_i x_j.
\end{equation}
The summation is conducted over the coordinates $(x_1, x_2)$ of the
pixels that lie within an aperture centered at the origin of the
coordinate system to which $(x_1, x_2)$ refer.  Following the work of
\cite{2006ApJ...639...64O}, we use an aperture of radius $R_{500}$
centered on the brightness peak.  We then compute $\eta$ from the
ratio of the non-zero elements that result from diagonalizing the
matrix $M$.  That is,
\begin{equation}
  \label{eq:diagonalize}
  D = U^T M U,
\end{equation}
where $U$ is a diagonalizing matrix for $M$, and  
\begin{equation}
  \label{eq:aratio}
  \eta = 
  \left\{ 
    \begin{array}{ll}
      \frac{D_{12}}{D_{21}}, & D_{12} \le D_{21} \\
      \frac{D_{21}}{D_{12}}, & D_{12} > D_{21}
    \end{array}
  \right\} \; \; .
\end{equation}
The axial ratio is therefore defined to be in the range $\eta \in [0,
1]$, with $\eta = 1$ for a circular cluster.  Of course there are
other choices for the origin of the coordinate system, besides using
the brightness peak.  For instance, in order to avoid misplaced
apertures yielding artificially low axial ratios for nearly circular
distributions, one could adjust the position of the aperture to seek a
maximum in $\eta$.  Doing this, we sometimes find that $\eta \approx
1$ even for non-circular clusters, as is evident in Figure
\ref{fig:badaxialratio}.  This figure depicts the surface-brightness
map of what appears to be a disturbed cluster, chosen from among those
in our sample that appear by eye to be the most unrelaxed. Yet, it
happens to have an axial ratio very close to 1 for an aperture placed
so as to maximize $\eta$.  This example demonstrates that, while the
axial ratio statistic may yield results consistent with a visual
interpretation of cluster substructure, it is also capable of
unexpected results for some clusters.

\begin{figure}[t]
  \centering
  \includegraphics[width=0.5\textwidth,trim=0mm 10mm 0mm 10mm]{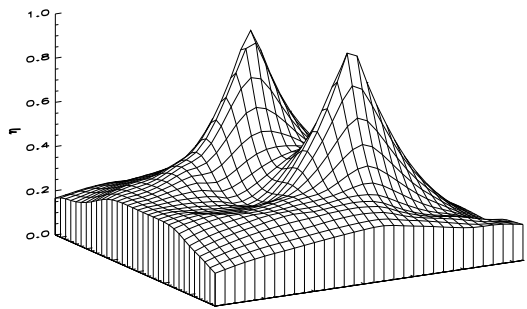}
  \caption{Surface of axial ratio $\eta$ as a two-dimensional function
    of the coordinates of the aperture center.  The axial ratio
    statistic appears to be ill-defined for this cluster.}
  \label{fig:badaxialratiosurface}
\end{figure}

\begin{figure}[t]
  \centering
  \includegraphics[width=0.5\textwidth,trim=0mm 10mm 0mm 10mm]{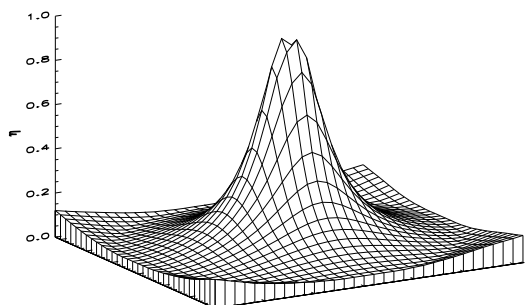}
  \caption{Abstract surface of axial ratio $\eta$ as a two-dimensional
    function of the coordinates of the aperture center.  This is a
    relaxed cluster, for which the axial ratio is better-defined.}
  \label{fig:goodaxialratiosurface}
\end{figure}

\begin{figure*}[t]
  \centering
  \includegraphics[width=1.0\textwidth,trim=5mm 0mm 0mm
  0mm,clip]{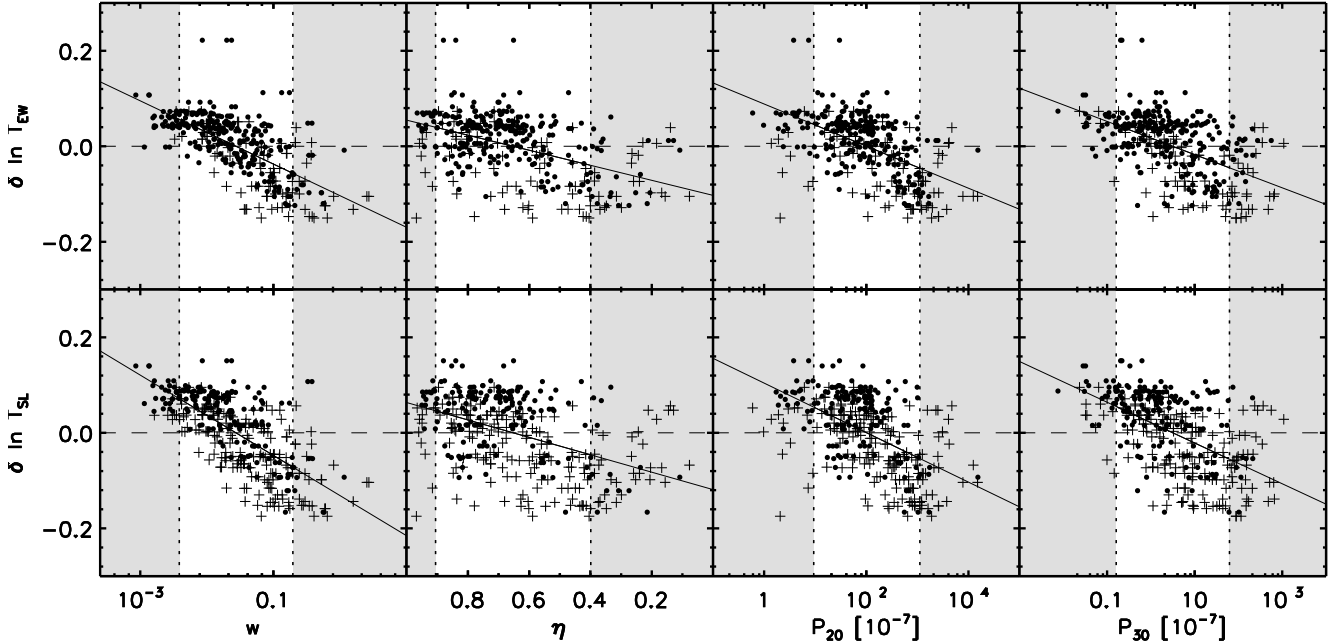}
  \caption{Relationship between a cluster's deviation $\delta \ln \tx$
    from the mean \mtx\ relationship and four measures of
    substructure: the centroid variation $w$, the axial ratio $\eta$,
    and the power ratios $P_{20}$ and $P_{30}$.  The lower panels are
    for spectroscopic-like temperature, and the upper ones are for
    emission-weighted temperature.  The light gray bands and vertical
    dashed lines show the extremes in substructure measurements, with
    80\% of our clusters falling in the central region between the
    bands.  The black filled circles correspond to clusters above 2
    keV, while the plus signs correspond to clusters below 2 keV.
    Finally, the solid lines indicate the best-fitting linear
    relationships between our substructure measures and temperature
    offset, for the above 2 keV sub-sample denoted by the black filled
    circles.}
  \label{fig:substruct_dev_temp}
\end{figure*}

To further illustrate this point, we have computed an axial ratio
value for this cluster for every possible choice of aperture
placement.  Apertures of radius $R_{500}$ were centered on each and
every pixel within the surface-brightness map, provided the aperture
so placed does not reach the edge of the map.  This procedure
generated an axial-ratio ``surface'' mapping all of the aperture
placements to a value of $\eta$.  Figure
\ref{fig:badaxialratiosurface} shows the axial-ratio surface for the
cluster in Figure \ref{fig:badaxialratio}.  For comparison purposes,
Figure \ref{fig:goodaxialratiosurface} presents an axial-ratio surface
map for a very symmetric, uniform, and apparently relaxed cluster, in
which the cluster's brightness peak reassuringly corresponds to the
aperture location that maximizes $\eta$.  In contrast, the presence of
two peaks in the axial-ratio surface for the asymmetric cluster shows
that $\eta$ can sometimes depend strongly on aperture placement.
Ideally, we would like to place the aperture on the ``center'' of this
cluster, but the center of an unrelaxed cluster can be difficult to
define, meaning that the axial ratio statistic may be likewise
ill-defined for such clusters.

\subsection{Power Ratio}
The power-ratio statistics
\citep{1995ApJ...452..522B,2006ApJ...639...64O} quantify substructure
by decomposing the surface-brightness image into a two-dimensional
multipole expansion, the terms of which are calculated from the
moments of the image, computed within an aperture of radius $\Rap$:
\begin{equation}
  \label{eq:a_m}
  a_m(\Rap) = \int_{R' \leq \Rap}\Sigma(\vec x')(R')^m \cos m \phi'\ d^2x'
\end{equation}
\begin{equation}
  \label{eq:b_m}
  b_m(\Rap) = \int_{R' \leq \Rap}\Sigma(\vec x')(R')^m \sin m \phi'\ d^2x'.
\end{equation}
The power in terms of order $m$ is then
\begin{equation}
  \label{eq:pm}
  P_m = \frac{(a_m^2 + b_m^2)}{2m^2\Rap^{2m}}.
\end{equation}
For $m = 0$, the power is given by
\begin{equation}
  \label{eq:p0}
  P_0 = [a_0 \ln(\Rap)]^2.
\end{equation}
The power ratios $P_{m0} \equiv P_m / P_0$ are then dimensionless
measures of substructure which have differing interpretations.  For
instance, $P_{10}$ quantifies the degree of balance about some origin
and can be used to find the image centroid, $P_{20}$ is related to the
ellipticity of the image, and $P_{30}$ is related to the triangularity
of the photon distribution.  As in the case of the axial ratio
computations, we set the aperture radius $\Rap$ equal to $R_{500}$.
The most appropriate place to center the aperture is at the set of
pixel coordinates that minimizes $P_{10}$, which we achieve using a
self-annealing algorithm.

\subsection{Centroid Variation}
The centroid variation statistic $w$ is a measure of the center shift,
or ``skewness'', of a two-dimensional photon distribution.  It is
measured for a cluster surface-brightness map in the following way.
For a set of surface-brightness levels one finds the centroids of the
corresponding isophotal contours and computes the variance in the
coordinates of those centroids, scaled to $R_{500}$.  Here we select
10 isophotes evenly spaced in $\log I_{\rm X}$ between the minimum and
maximum of $I_{\rm X}$ within an aperture of radius $R_{500}$ centered
on the brightness peak, so as to adapt to the full dynamic range of
surface brightness for different clusters.  We employed this adaptive
scheme because using one set of isophotes for all clusters tended to
ignore important substructure in less massive clusters when they had
surface brightness substructure inside $R_{500}$ but outside of the
lowest isophote.

\subsection{Substructure and Scaling Relationships}

Using the quantitative measures of substructure described in the
previous section, we can test the significance of the relationship
between substructure and temperature offset hinted at in Figure
\ref{fig:cluster_gallery_extremes}.  We begin by treating four of our
substructure statistics---centroid variation $w$, axial ratio $\eta$,
and power ratios $P_{20}$ and $P_{30}$---as different imperfect
measurements of an intrinsic degree of substructure $S$.  Figure
\ref{fig:substruct_dev_temp} shows the relationship between
substructure and a cluster's deviation $\delta \ln \tx$ from the mean
\mtx\ relation for each substructure measure.  In each case we present
results for both the spectroscopic-like temperature $\tsl$ and the
emission-weighted temperature $\tew$.  Note that centroid variation
$w$ and the power ratios $P_{20}$ and $P_{30}$ have large dynamic
ranges, whereas the axial ratio $\eta$ is always of order unity.  We
therefore attempt to fit the relationships between $\delta \ln \tx$
and the different substructure measures with the following forms:
\begin{equation}
  \delta \ln \tx = \left\{
  \begin{array}{l}
    A w^{\alpha} \\
    B + \beta\eta \\
    C P_{20}^{\gamma} \\
    D P_{30}^{\lambda}
   \end{array} \right.
\end{equation}
To visually indicate where the bulk of our substructure measures lie,
Figure \ref{fig:substruct_dev_temp} has light gray bands covering the
extremes, so that 80\% of our sample clusters have substructure
measures lying between the extremes.  The power ratios in our study
generally span two decades (in units of $10^{-7}$), from $\sim$2---300
for $P_{20}$ and from $\sim$0.01---10 for $P_{30}$.  These ranges are
consistent with those of \cite{1995ApJ...452..522B},
\cite{2006ApJ...639...64O}, and \cite{2007arXiv0708.1518J}.  The
measurements of axial ratio in our sample, with 80\% of clusters
having $\eta$$\sim$0.4---0.95, cover a slightly wider range than do
the simulated clusters of \cite{2006ApJ...639...64O}.  Finally, our
measurements of centroid variation, with 80\% of clusters having
$w[R_{500}]$$\sim$0.01---0.1, are again similar to those of
\cite{2006ApJ...639...64O}.

As denoted in Figure \ref{fig:substruct_dev_temp} by black filled
circles, the systems with $\tx$ above 2 keV occupy a slightly narrower
range of substructure values than the systems below 2 keV, which are
denoted by plus signs.  For the axial ratio and the power ratios, the
variance is 15 to 25 percent larger among the low-temperature systems
when compared to the systems with $\tx >$ 2 keV.  For centroid
variation the variance among the low-temperature systems is
approximately the same as it is among the high-temperature systems.
However, it is not clear that there is a significant correlation
between substructure and mass, since the mean substructure values are
generally very similar between the low-temperature and
high-temperature subsamples.  The mean value of the power ratio
$P_{30}$ is significantly larger for the low-temperature subsample,
however this measure also has the weakest correlation with offsets
from the mean \mtx\ relation.

\begin{figure}[t]
  \centering
  \includegraphics[width=0.5\textwidth,trim=5mm 0mm 0mm
  0mm,clip]{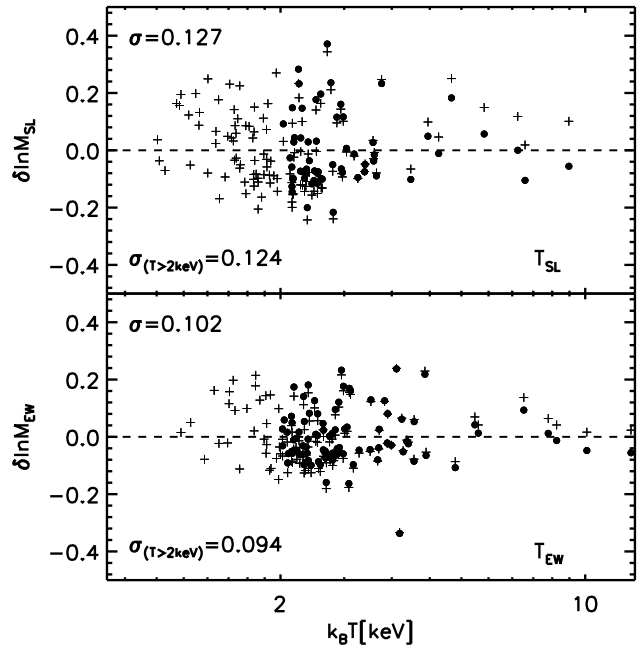}
  \caption{Comparison of mass offset $\delta \ln M$ between true mass
    and predicted mass, based on the $M(\tx)$ relation.  Plus signs
    indicate residuals for masses estimated from the $M$-$T_{\rm X}$
    relation derived from all 121 clusters, while black filled circles
    are for masses estimated from the $M$-$T_{\rm X}$ relation for
    clusters above 2 keV.  Upper panels are for spectroscopic-like
    temperature and lower panels are for emission-weighted
    temperature.  The standard deviations $\sigma$ in the residuals
    are given in each plot.}
  \label{fig:resid_lowt}
\end{figure}

To test whether the low-mass clusters in our sample significantly
boost the overall scatter in the \mtx\ relation, we perform a cut at 2
keV and fit this relation both to the whole sample and to the
sub-sample above 2 keV.  Figure \ref{fig:resid_lowt} shows the
residuals in mass, actual minus predicted, where the predicted mass
derives only from the \mtx\ relation.  The plus signs indicate
clusters whose mass is predicted from an \mtx\ relation derived from
all 121 clusters.  The black filled circles indicate clusters that are
above 2 keV in X-ray temperature, with the mass estimated using the
sub-sample \mtx\ relation.  There is a negligible reduction in
scatter, from 0.127 to 0.124 for $\tsl$ and from 0.102 to 0.094 for
$\tew$, suggesting that at best only a modest improvement is found in
our sample if we remove the low-mass systems.  In order to test the
degree to which incorporating substructure measures adds to this
modest improvement, when we compare mass estimates derived using
substructure to those derived only from the \mtx\ relation, we focus
on clusters above 2 keV in the rest of our analysis.

Figure \ref{fig:substruct_dev_temp} shows that for our simulated
clusters, a greater amount of measured substructure tends to be
associated with ``cooler'' clusters while less substructure tends to
be associated with ``hotter'' clusters.  Also, the centroid variations
$w$ are more highly-correlated with $\delta \ln \tx$ than are the
other substructure parameters.  We interpret this to mean that the
centroid variation is a better predictor of the offset in the \mtx\
relationship than are the power ratios and the axial ratio, though all
four measures appear to be related to the temperature offset.  Again,
in this figure we denote systems above 2 keV by black filled circles,
and systems below 2 keV by plus signs.

\begin{figure*}[t]
  \centering
  \includegraphics[width=1.0\textwidth,trim=5mm 0mm 0mm
  0mm,clip]{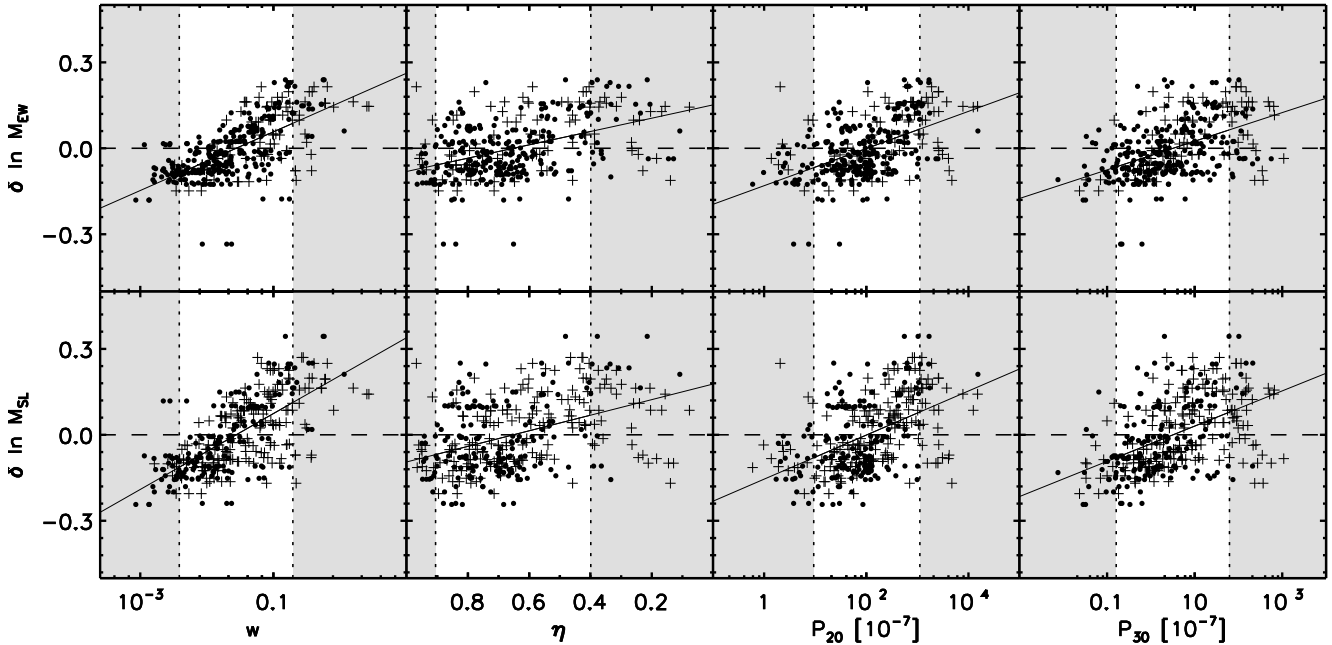}
  \caption{Relationship between substructure and mass offset $\delta
    \ln M (\tx)$ from the mean \mtx\ relationship for the same
    substructure measures as in Figure \ref{fig:substruct_dev_temp}.
    The lower panels are for $\tx = \tsl$, and the upper panels are
    for $\tx = \tew$.  As in Figure \ref{fig:substruct_dev_temp} the
    solid lines indicate the best-fitting linear relationships to the
    above 2 keV sub-sample denoted by black filled circles, and the
    gray bands and vertical dashed lines mark the extremes in
    substructure between which 80\% of our clusters lie.  Also as in
    Figure \ref{fig:substruct_dev_temp}, the plus signs correspond to
    systems below 2 keV., while the black filled circles correspond to
    systems above 2 keV.}
  \label{fig:substruct_dev}
\end{figure*}

Correlations between substructure and $\delta \ln \tx$ can potentially
be exploited to improve on mass estimates of real clusters derived
from the mass-temperature relation.  Instead of computing the
temperature offset at fixed mass, we can determine a
substructure-dependent mass offset at fixed temperature and then apply
it as a correction to the predicted mass $M_{\rm pred}(\tx)$ one would
derive from the mean \mtx\ relation alone.  To assess the prospects
for such a correction, based on this sample of simulated clusters, we
first define the mass offset from the mean mass-temperature relation
to be
\begin{equation}
  \delta \ln M (\tx) = \ln \left[ \frac {M} {M_{\rm pred}(\tx)} \right] \; \; ,
  \label{eq:deltaM}
\end{equation}
where $M$ is the cluster's actual mass, and examine the correlations
between substructure measures and $\delta \ln M$.  Figure
\ref{fig:substruct_dev} shows the results.  These plots show mass
predictions from both the $M$-$\tsl$ relation and the $M$-$\tew$
relation.  Consistent with our analysis of $\delta \ln \tx$, the
centroid variation $w$ appears to be a more effective predictor of the
mass offset $\delta \ln M (\tx)$.  Nonetheless, all four measures of
substructure appear to be correlated with mass offset.

\begin{figure*}[t]
  \centering
  \includegraphics[width=1.0\textwidth,trim=5mm 0mm 0mm
  0mm,clip]{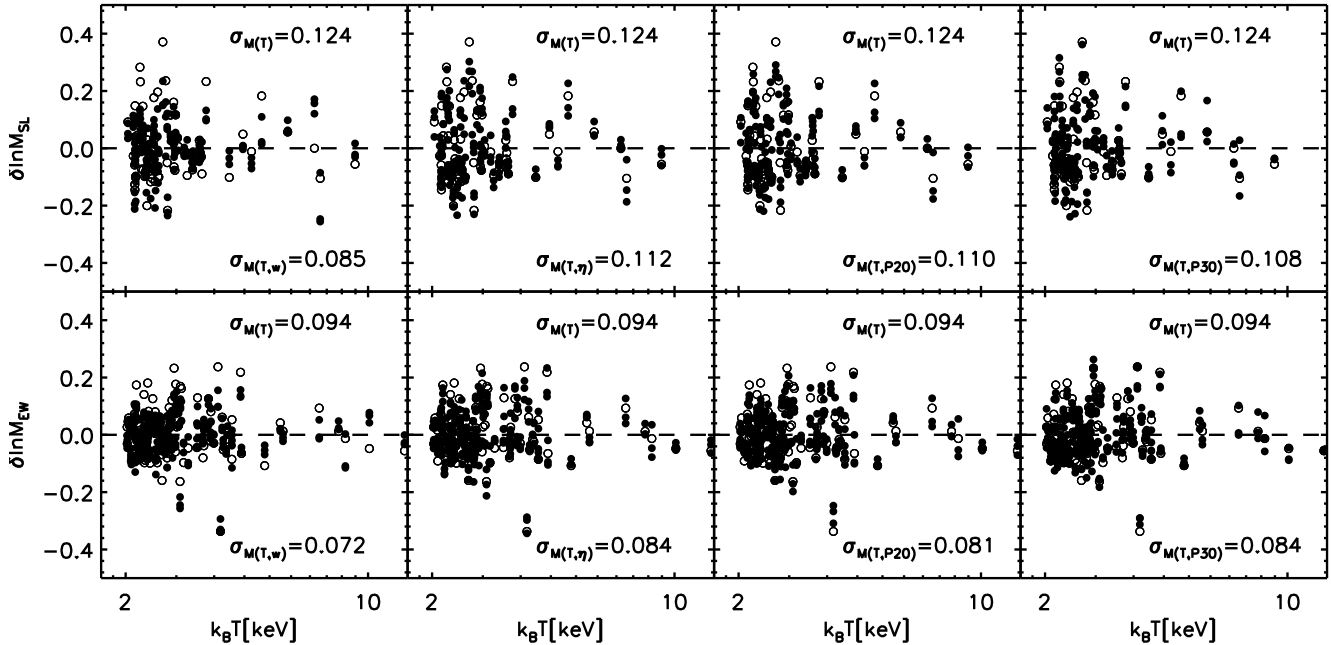}
  \caption{Comparison of mass offset $\delta \ln M$ between true mass
    and predicted mass, based on the $M(\tx)$ relation (open circles)
    and the $M(\tx,S)$ relation (filled circles).  Upper panels are
    for $\tsl$ and lower panels are for $\tew$.  The standard
    deviations $\sigma$ in the residuals are given in each plot.}
  \label{fig:resid2}
\end{figure*}

In order to incorporate a substructure correction into the
mass-temperature relation, we perform a multiple regression, fitting
our simulated clusters' mass, temperature, and substructure data to
the form,
\begin{equation}
  \log M_{\rm pred} (\tx,S) = \log M_0 + \alpha \log \tx + \beta S ,
  \label{eq:mass-temp-subst}
\end{equation}
where $S$ represents one of the following substructure measures:
$\eta$, $\log w$, $\log P_{20}$, or $\log P_{30}$.  This fit gives us
a substructure-corrected mass prediction $M_{\rm pred} (\tx, S)$ for
each substructure measure, and we can assess the effectiveness of that
correction by measuring the dispersion of the substructure-corrected
mass offset
\begin{equation}
  \delta \ln M (\tx,S) = \ln \left[ \frac {M} {M_{\rm pred}(\tx,S)} \right] \; \; ,
  \label{eq:deltaMTS}
\end{equation}
between the revised prediction and the true cluster mass.

Figure \ref{fig:resid2} shows the results of that test.  Open circles
in each panel indicate mass offsets $\delta \ln M(\tx)$ without
substructure corrections, which have a standard deviation
$\sigma_{M(T)}$.  Filled circles indicate mass offsets $\delta \ln
M(\tx,S)$ with substructure corrections, which have a standard
deviation $\sigma_{M(T,S)}$.  The upper set of panels shows results for
$\tsl$, and the lower set is for $\tew$.  In each case, incorporating
a substructure correction to the mass-temperature relation reduces the
scatter, yielding more accurate mass estimates.  The centroid
variation corrections are the most effective, reducing the scatter in
mass from 0.124 to 0.085 in the $M$-$\tsl$ relation and from 0.094 to
0.072 in the $M$-$\tew$ relation, though admittedly this is again a
modest improvement.  Although non-negligible structure correlates
significantly with offsets in the \mtx\ plane, apparently it does so
with substantial scatter.  This scatter may be partly due to
projection effects, in which line-of-sight mergers are discounted by
the measures of substructure and may dilute their corrective power
\citep{2007arXiv0708.1518J}.

\begin{figure*}[t]
  \centering
  \includegraphics[width=1.0\textwidth,trim=5mm 30mm 0mm
  0mm,clip]{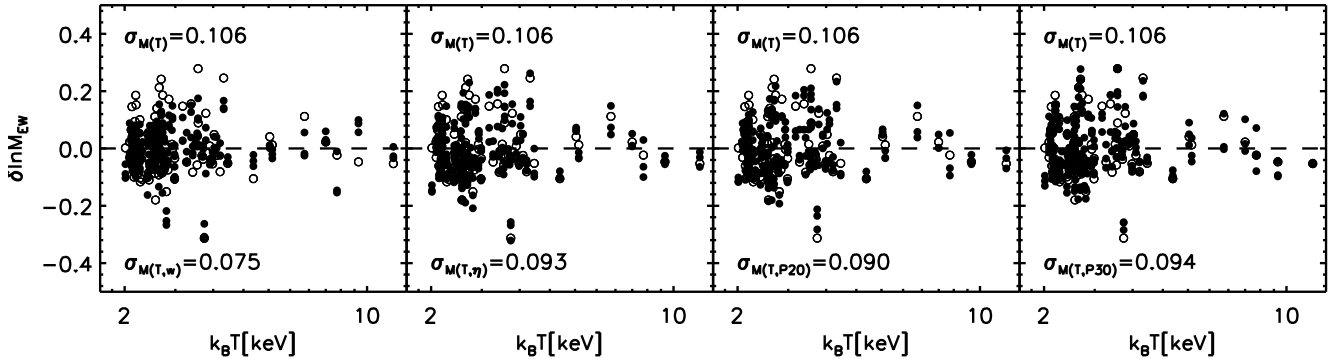}
  \caption{Comparison of mass offset $\delta \ln M$ between true mass
    and predicted mass, based on the $M(\tx)$ relation (open circles)
    and the $M(\tx,S)$ relation (filled circles).  These results are
    for are for $\tew$, with the central 0.15 $R_{500}$ region removed
    both from the average temperature and from the substructure
    measures.  The standard deviations $\sigma$ in the residuals are
    given in each plot.}
  \label{fig:resid3}
\end{figure*}

Lastly, Figure \ref{fig:resid3} shows the results for a similar
analysis to that of Figure \ref{fig:resid2}, except that in this case
we have excised a region of radius 0.15$R_{500}$ around the center of
each cluster and recomputed $\tew$.  We do this to test whether offset
in temperature, whose correlation with substructure is the basis of
our correction scheme, stems from a potentially unrealistic feature,
which is that the cores of many real clusters have temperature
profiles that decline at larger radii than occurs in simulated
clusters.  As in Figure \ref{fig:resid2}, we restrict our analysis to
clusters above 2 keV.  After doing this test, for $\tew$ excising the
core actually increases the scatter in \mtx\ from 0.094 to 0.106.  It
may be that by removing the bright central region, the average
temperature becomes more sensitive to structure outside the core.
Also, this figure shows that the effect of incorporating substructure
measurements into the mass-estimates is still present.  The scatter is
reduced to 0.075 for $w$, 0.093 for $\eta$, 0.090 for $P_{20}$ and
0.094 for $P_{30}$.  Figure \ref{fig:resid3} summarizes the results of
this test, which support the conclusion that the reduction in scatter
we realize using substructure is a real effect and not an artifact of
known defects in the simulations.

\subsection{Comparisons with Other Substructure Studies}

\cite{2006ApJ...639...64O} examined the relationship between galaxy
cluster substructure and X-ray scaling relationships, including the
\mtx\ relation, using both a flux-limited sample of nearby clusters
and a sample of simulated clusters, and found a greater amount scatter
among the more relaxed clusters in their observed sample.  Contrasting
that result they also found a greater amount of scatter among the more
disrupted clusters in their simulation sample, though they
characterize the evidence for this second result to be weak.  Finally,
they see no evidence in either sample for more disrupted clusters to
be below the mean, and the more relaxed clusters to be above.  One
difference between our study and theirs is the presence of radiative
cooling and supernova feedback in the simulation that produced our
cluster sample.  Also, the focus of our work is different from theirs
in that we concentrate on the degree of correlation between the amount
of substructure and the size and direction of the offset from the mean
relation.  We do find significant evidence of this correlation, such
that relaxed clusters are hotter than expected given their mass.  We
also test, as best we can given our simulation sample, the hypothesis
that substructure can be used to improve mass estimates derived from
the ICM X-ray temperature.  It is possible that our detection of a
correlation between substructure and temperature offset arises from
the additional physics in our simulated clusters, since when radiative
cooling is included, cool lumps may be better preserved than in
simulations that don't include cooling.

Our results are in agreement with \cite{2006NewA...12...71V}, who
examined substructure in clusters simulated with cooling and feedback
and found that unrelaxed clusters, identified with a larger power
ratio $P_{30}$, have spectral-fit temperatures biased low relative to
the mass-weighted temperatures.  This trend aligns with our finding
that the spectroscopic-like temperature $\tsl$ is lower than the
best-fit temperature at fixed mass for clusters with larger power
ratios $P_{20}$ and $P_{30}$.  However, \cite{2006NewA...12...71V} did
not investigate the effectiveness of substructure measures in reducing
scatter in the mass-temperature relation.

Our results are also in agreement with some of the results of
\cite{2007arXiv0708.1518J}, who have recently investigated
correlations between substructure and offsets in mass predictions in
simulated clusters.  They found that measuring cluster structure is an
effective way to correct masses estimated using the assumption of
hydrostatic equilibrium, which tend to be underestimates.  Our
findings support these results, given that we find substructure can be
used to correct masses estimated directly from the \mtx\ relationship.
There also are differences between our findings and theirs.  They
report that the \mtx\ relation for their simulation sample shows no
dependence on structure, whereas the clusters in our sample exhibit
offsets that correlate with the degree of substructure.  One
possibility is that these differences stem from differences in the
simulations' feedback mechanisms.  Another possibility is that some of
the offset we observe derives from enhanced temperatures in
simulations with radiative cooling.  As we describe in section 3, we
perform a test in which we estimate $\tew$\ using projected
surface-brightness and temperature maps, in order to remove the core
regions from our analysis, but this may be less effective than
properly excising the cores in the simulations, as
\cite{2007arXiv0708.1518J} have done.

\cite{2006ApJ...650..128K} also looked at the relationship that
cluster structure has to the \mtx\ relation in simulated clusters, to
show that the sensitivity of mass proxies $Y_X$ and $Y_{SZ}$ to
substructure is not very strong.  They divided their sample into
unrelaxed and relaxed subsamples, based on the presence or absence of
multiple peaks in the surface-brightness maps of clusters, and found
the normalization of the \mtx\ relation to be biased to cooler
temperatures for the unrelaxed systems.  Other workers also have
looked at the relationship between the \mtx\ relation and
substructure, as reflected in the X-ray spectral properties.
\cite{2001ApJ...546..100M} have examined the ratio of X-ray
spectral-fit temperatures in hard and full bandpasses for an ensemble
of simulated clusters, and found it to be a way of quantifying the
dynamical state of a cluster.  We consider our approach of using
surface-brightness morphology information to be complementary to
theirs.  More recently, \cite{2007MNRAS.377..317K} performed an
interesting analysis on another large-volume simulation sample, using
as substructure metrics the centroid variation and measures of
concentration to report evolution in the luminosity-temperature
relationship.  Specifically, they report that the more irregular
clusters in their sample lie above the mean \mtx\ relation (i.e., they
are cooler than average), for the spectroscopic-like temperature
$\tsl$.

\section{Summary}

Using a sample of galaxy clusters simulated with cooling and feedback,
we investigated three substructure statistics and their correlations
with temperature and mass offsets from mean scaling relations in the
\mtx\ plane.  First, we showed that the substructure statistics $w$,
$\eta$, $P_{20}$ and $P_{30}$ all correlate significantly with $\delta
\ln \tx$, though with non-negligible scatter.  In all cases this
scatter is larger for $\delta \ln$ $\tsl$ than it is for $\delta \ln$
$\tew$.  Next, we considered the possibility that \mtx\ scatter is
driven by low-mass clusters.  We tested the degree to which scatter
can be reduced by filtering out these systems.  This consisted of
performing a cut at 2 keV, for which we saw that it yielded a modest
improvement in mass estimates.  To see whether incorporating
substructure could refine these mass estimates, we first showed that
$w$, $\eta$, $P_{20}$, and $P_{30}$ correlate significantly with the
difference $\delta \ln M$ between masses predicted from the mean
$M(\tx)$ relation and the true cluster masses, with non-negligible
scatter that again is less for $M(\tew)$ than it is for $M(\tsl)$.
Then we adopted a full three-parameter model, $M$-$\tx$-$S$, which
includes substructure information $S$ estimated using $w$, $\eta$,
$P_{20}$, and $P_{30}$. Scatter about the basic two-parameter
$M$-$\tew$ relation was 0.094.  Including substructure as a third
parameter reduced the scatter to 0.072 for centroid variation, 0.084
for axial ratio, 0.081 for $P_{20}$, and 0.084 for $P_{30}$.  Scatter
about the basic two-parameter $M$-$\tsl$ relation was 0.124, and
including substructure as a third parameter reduced the scatter to
0.085 for centroid variation, 0.112 for axial ratio, 0.110 for
$P_{20}$, and 0.108 for $P_{30}$.  As one last test, and to increase
our confidence that our substructure measures are not relying on
potentially non-physical core structure in the simulations, we also
repeated the comparison of mass-estimates for $\tew$, with the core
regions of the clusters excised.  First, removing the core slightly
increased the scatter in \mtx\, possibly by making the average
temperature more sensitive to structure outside the core.  Second,
even with the cores removed the improvement in mass-estimates obtained
using substructure information remains.  Based on these results, it
appears that centroid variation is the best substructure statistic to
use when including a substructure correction in the $M$-$\tew$
relation.  However, the correlations we have found in this sample of
simulated clusters might not hold in samples of real clusters, because
relaxed clusters in the real universe tend to have cooler cores than
our simulated clusters do.

\acknowledgments The authors wish to thank Stefano Borgani for
contributing the simulation data on which this project was based and
for his helpful comments on the manuscript.  This work was supported
by NASA through grants NNG04GI89G and NNG05GD82G, through Chandra
theory grant TM8-9010X, and through Chandra archive grant
SAOAR5-6016X.

\end{document}